\documentclass[preprint,showpacs,preprintnumbers,amsmath,amssymb]{revtex4-1}

\usepackage[final]{graphicx}
\usepackage{hyperref}
\usepackage{subfigure}
\usepackage{enumerate}
\usepackage{color}

\usepackage{dcolumn}% Align table columns on decimal point
\usepackage{bm}% bold math
\usepackage{ulem}
\usepackage{braket}
\usepackage{epstopdf}
\usepackage{amsmath}
\usepackage[percent]{overpic}
\usepackage{adjustbox}
\usepackage{booktabs}
\usepackage{hyperref}
\usepackage{cancel}

\usepackage{ulem}

\usepackage{array}
\newcolumntype{C}[1]{>{\centering\let\newline\\\arraybackslash\hspace{0pt}}m{#1}}

\begin{document}
\title{Complex Magnetic Order in a Decorated Spin Chain System Rb$_2$Mn$_3$(MoO$_4$)$_3$(OH)$_2$}

\author{Yaohua~Liu} \email[]{liuyh@ornl.gov. This manuscript has been authored by UT-Battelle, LLC under Contract No. DE-AC05-00OR22725 with the U.S. Department of Energy. The United States Government retains and the publisher, by accepting the article for publication, acknowledges that the United States Government retains a non-exclusive, paid-up, irrevocable, world- wide license to publish or reproduce the published form of this manuscript, or allow others to do so, for United States Government purposes. The Department of Energy will provide public access to these results of federally sponsored research in accordance with the DOE Public Access Plan(http://energy.gov/downloads/doe-public-access-plan).}
\affiliation{Neutron Science Division, Oak Ridge National Laboratory, Oak Ridge, Tennessee 37831, USA}

\author{Liurukara~D.~Sanjeewa}
\affiliation{Materials Science and Technology Division, Oak Ridge National Laboratory, Oak Ridge, Tennessee 37831, USA}

\author{V.~Ovidiu~Garlea}
\affiliation{Neutron Science Division, Oak Ridge National Laboratory, Oak Ridge, Tennessee 37831, USA}

\author{Tiffany~M.~Smith~Pellizzeri}
\affiliation{Department of Chemistry and Center for Optical Materials Science and Engineering Technologies (COMSET), Clemson University, Clemson, South Carolina 29634,USA }
\affiliation{Department of Chemistry and Biochemistry, Eastern Illinois University, Charleston, Illinois 61920, USA}

\author{Joseph~W.~Kolis}   
\affiliation{Department of Chemistry and Center for Optical Materials Science and Engineering Technologies (COMSET), Clemson University, Clemson, South Carolina 29634,USA }

\author{Athena~S.~Sefat}
\affiliation{Materials Science and Technology Division, Oak Ridge National Laboratory, Oak Ridge, Tennessee 37831, USA}

\date{\today}

\begin{abstract}
Macroscopic magnetic properties and microscopic magnetic structure of Rb$_2$Mn$_3$(MoO$_4$)$_3$(OH)$_2$  (space group $Pnma$) are investigated by magnetization, heat capacity and single-crystal neutron diffraction measurements. The compound's crystal structure contains bond-alternating [Mn$_3$O$_{11}$]$^{\infty}$ chains along the $b$-axis, formed by isosceles triangles of Mn ions occupying two crystallographically nonequivalent sites (Mn1 site on the base and Mn2 site on the vertex). These chains are only weakly linked to each other by nonmagnetic oxyanions. Both SQUID magnetometry and neutron diffraction experiments show two successive magnetic transitions as a function of temperature.  On cooling, it transitions from a paramagnetic phase into an incommensurate phase below 4.5~K with a magnetic wavevector near ${\bf k}_{1} = (0,~0.46,~0)$. An additional commensurate antiferromagnetically ordered component arises with ${\bf k}_{2} = (0,~0,~0)$, forming  a complex magnetic structure below 3.5~K with two different propagation vectors of different stars. On further cooling, the incommensurate wavevector undergoes a lock-in transition below 2.3~K.  The experimental results suggest that the magnetic superspace group is $Pnma.1'(0b0)s0ss$ for the single-${\bf k}$ incommensurate phase and is $Pn'ma(0b0)00s$ for the 2-${\bf k}$ magnetic phase. We propose a simplified magnetic structure model taking into account the major ordered contributions,  where the commensurate ${\bf k}_{2}$ defines the ordering of the $c$-axis component of Mn1 magnetic moment, while the incommensurate ${\bf k}_{1}$ describes the ordering of the $ab$-plane components of both Mn1 and Mn2 moments into elliptical cycloids. 

\end{abstract}
% spin arrangement in magnetically ordered materials; antiferromagnetics, and neutron diffraction
\maketitle

\section{\label{sec:intro}Introduction}
Geometrically frustrated magnetic systems host many interesting electronic and magnetic phenomena and have attracted enduring research efforts~\cite{ramirez1994strongly, gardner2011geometrically, nisoli2013colloquium}. Certain oxyanion materials containing magnetic transition-metal oxide motifs that are magnetically separated from each other by nonmagnetic ligands, e.g., closed-shell ion clusters [AsO$_4$]$^{3-}$, [MoO$_4$]$^{2-}$ and [VO$_4$]$^{3-}$, have become a fertile ground to investigate emergent quantum phenomena in low-dimensional frustrated magnetic systems~\cite{hwu1998structurally, garlea2019exotic, sanjeewa2019magnetic}. Among these materials, delta spin chain (a.k.a. sawtooth chain) systems are of particular interest because of geometrical frustrations~\cite{sen1996quantum,van2001fine, garlea2014complex,nakamura1996elementary, blundell2003quantum, kikuchi2011spin, lau2006magnetic}. For example, recent studies on Rb$_2$Fe$_2$O(AsO$_4$)$_2$ have found complex magnetic behaviors that originate from strongly frustrated interactions within the sawtooth chains and relatively weak coupling between them~\cite{garlea2014complex}. 

While Rb$_2$Fe$_2$O(AsO$_4$)$_2$ has sawteeth on opposite sides of the spin chain, the newly discovered compound Rb$_2$Mn$_3$(MoO$_4$)$_3$(OH)$_2$ is a bond-alternating chain system~\cite{pellizzeri2019alkali}, where all the sawteeth of each spin chain are sitting on the same side with no shared magnetic ions between adjacent sawteeth. It looks like a spin chain (Mn1 sites) decorated with magnetic-ion pendants (Mn2 sites) (see Fig.~\ref{Fig:sample}). Here we report the magnetic properties and magnetic structure of the latter compound, which contains three dissimilar exchange interactions between nearest neighbors inside each individual spin delta chain. From magnetic susceptibility and heat capacity measurements as well as neutron diffraction experiments, we have found two successive magnetic transitions in Rb$_2$Mn$_3$(MoO$_4$)$_3$(OH)$_2$ on cooling. It enters into an incommensurate phase below 4.5~K with a magnetic wavevector near ${\bf k}_{1} = (0,~0.46,~0)$. A second antiferromagnetically ordered component shows up with a commensurate wavevector ${\bf k}_{2} = (0,~0,~0)$, which coexists with the incommensurate component below 3.5~K, forming a complex magnetic structure with two different propagation vectors of different stars. On further cooling, the incommensurate wavevector undergoes a lock-in transition below 2.3~K.  Our experimental results show that the magnetic superspace group is $Pnma.1'(0b0)s0ss$ (in standard setting: $Pbnm1'(00g)s00s$ (62.1.9.4m442.2)) for the single-$k$ incommensurate phase and is further lowered down to $Pn'ma(0b0)00s$ (in standard setting: $Pbn'm(00g)s00$ (62.1.9.4m443.1)) for the 2-$k$ magnetic phase. For the major ordered contributions, the commensurate ${\bf k}_{2}$ defines the ordering of the $c$-axis component of Mn1 magnetic moment, while the incommensurate ${\bf k}_{1}$ comes from ordering of the $ab$-plane components of both Mn1 and Mn2 moments into elliptical cycloids. To our best knowledge, this is the first realization of such a bond-alternating delta chain. The experimental results will be the benchmark for future theoretical investigations on the complex magnetic properties of this new decorated spin-chain system.

\section{Samples and Experiments}
\begin{figure}[tb]
	\centering
		\includegraphics[width=0.48\textwidth]{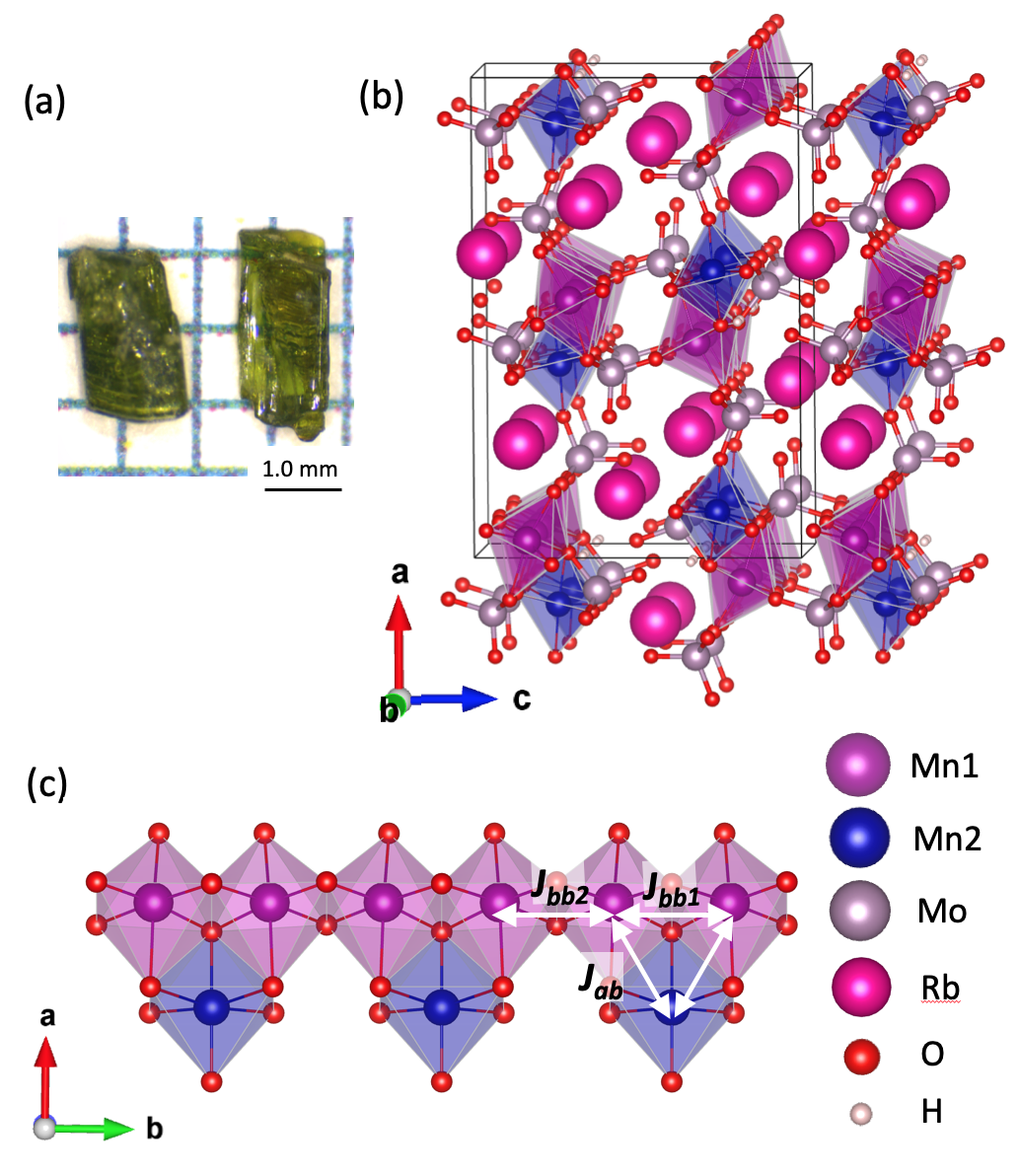} 
		\caption{ \label{Fig:sample} (Color online) Typical Rb$_2$Mn$_3$(MoO$_4$)$_3$(OH)$_2$ crystals used in this study have a dark-green color and a prismatic shape with an approximate size of $0.5 \times 1.0 \times 2.5$~mm$^3$. (b) The extended crystal structure shows the structural confinement of the Mn chains. (c) Mn1 forms the chains, which are decorated with Mn2 pendants. Mn2O$_6$ octahedra share edges between two Mn1O$_6$ octahedra to form isosceles triangles. The quasi-one-dimensional spin chain looks like 
		sawteeth repeating along the $b$-axis with alternating bond lengths.}
\end{figure}

\begingroup
\renewcommand{\arraystretch}{0.64}
\begin{table}[tb]
\small
\centering
\begin{tabular}{c|ccc|c|c}
  \hline
  \hline
  Atom  &        $x$     &      $y$     &      $z$       &   $U_{eq}$    &  Wyck.\\
  \hline
  Rb1   &   0.36007(5) & -1/4        & -0.04187(7)  & 0.0178(2)   &  $4c$\\
  Rb2   &   0.25567(5) & 3/4         & 0.65004(9)   & 0.0297(3)   &  $4c$\\
  \hline
  Mn1   &   0.55100(5) & 0.4975(2)   & 0.31663(7)   & 0.0094(2)   &  $8d$\\
  Mn2   &   0.39638(7) & 3/4         & 0.2542(1)    & 0.0091(3)   &  $4c$\\
  \hline
  Mo1   &   0.44079(4) & 1/4         & 0.11391(6)   & 0.0061(2)   &  $4c$\\
  Mo2   &   0.60677(4) & 3/4         & 0.56579(6)   & 0.0076(2)   &  $4c$\\
  Mo3   &   0.19834(4) & 3/4         & 0.31418(6)   & 0.0087(2)   &  $4c$\\
  \hline
  O1    &   0.5179(3)  & 1/4         & 0.1999(5)    & 0.008(1)    &  $4c$\\
  O2    &   0.3884(2)  & 0.0199(7)   & 0.1393(3)    & 0.0117(10)  &  $8d$\\
  O3    &   0.4712(3)  & 1/4         & -0.0142(5)   & 0.019(2)    &  $4c$\\
  O4    &   0.5785(4)  & 3/4         & 0.4305(5)    & 0.016(2)    &  $4c$\\
  O5    &   0.5730(2)  & 0.9838(7)   & 0.6288(3)    & 0.016(1)    &  $8d$\\
  O6    &   0.6984(4)  & 3/4         & 0.5731(7)    & 0.034(2)    &  $4c$\\
  O7    &   0.1526(2)  & 0.5207(7)   & 0.2679(4)    & 0.015(1)    &  $8d$\\
  O8    &   0.2872(3)  & 3/4         & 0.2652(6)    & 0.028(2)    &  $4c$\\
  O9    &   0.1976(4)  & 3/4         & 0.4507(6)    & 0.030(2)    &  $4c$\\
  O10   &   0.5078(3)  & 3/4         & 0.2307(5)    & 0.008(1)    &  $4c$\\
  O11   &   0.5811(4)  & 1/4         & 0.4118(5)    & 0.017(2)    &  $4c$\\
  \hline
  H10   &   0.499(4)   & 0.84(1)     & 0.176(4)     & 0.05 (fixed) &  $8d$\\
 \hline
 \end{tabular}
\caption{Refined structural parameters of Rb$_2$Mn$_3$(MoO$_4$)$_3$(OH)$_2$ from room-temperature single-crystal x-ray data~\cite{pellizzeri2019alkali}. The data were refined in the orthorhombic space group $Pnma$, and the refined lattice parameters at 300~K are $a = 18.3294(6)$~\AA, $b=6.2474(2)$~\AA~and $c=12.4969(4)$~\AA.}
\label{Tab:SXD}
\end{table}
\endgroup

Rb$_2$Mn$_3$(MoO$_4$)$_3$(OH)$_2$ single crystals have been grown by the high-temperature high-pressure hydrothermal method~\cite{pellizzeri2019alkali}. Figure~\ref{Fig:sample}(a) shows the picture of a couple of single crystals, which form prismatic dark green crystals with typical sizes of $0.5 \times 1.0 \times 2.5$~mm$^3$, with the long edge being closely along the $b$-axis. Table~\ref{Tab:SXD} shows the structural parameters determined from single crystal x-ray diffraction data collected at room temperature, including the atomic coordinates and the displacement parameters. Detailed descriptions of the synthesis and the structure have been published~\cite{pellizzeri2019alkali}. There are two crystallographically different Mn sites, Mn1 at a $8d$ position ($0.551,~0.498,~0.317$), and Mn2 at a $4c$ Wyckoff position ($0.397,~3/4,~0.254$), respectively. Mn1 forms chains running along the $b$-axis and MnO$_6$ octahedra of Mn1 and Mn2 share edges to form triangles. As illustrated in Fig.~\ref{Fig:sample}, these delta chains are bridged each other via MoO$_4$ units along the $a$- and $c$-axes to form a three-dimensional network structure. Due to the nonmagnetic nature of the Mo$^{6+}$ cation, the magnetic interactions between adjacent [Mn$_{3}$O$_{11}$]$^{\infty}$ chains via the Mn-O-Mo-O-Mn connection are expected to be much weaker than the intra-chain interactions. The Mn-O bond lengths and Mn-O-Mn bond angles are summarized in Tab.~\ref{Tab:bond}.  There are three different nearest-neighbor super-exchange interactions within a spin chain, including Mn1-Mn1 pairs within a sawtooth ($J_{bb1}$), Mn1-Mn1 pairs between adjacent sawteeth ($J_{bb2}$), and Mn1-Mn2 pairs between the base-vertex ($J_{bv}$) of a sawtooth. 

\begingroup
\renewcommand{\arraystretch}{0.64}
\begin{table}[tb]
\small
\centering
\begin{tabular}{p{0.4\textwidth}  p{0.4\textwidth}}
  \hline
  \hline
  Mn1-O1: 2.2315(5)~\AA & Mn1-O4: 2.2203(5)~\AA \\
  Mn1-O5: 2.408(4)~\AA & Mn1-O5: 2.176(5)~\AA \\
  Mn1-O10: 2.083(5)~\AA & Mn1-O11: 2.044(5)~\AA \\
  \hline
  Mn2-O2: 2.238(5)~\AA & Mn2-O5: 2.304(5)~\AA \\
  Mn2-O8: 2.034(6)~\AA & Mn2-O10: 2.092(6)~\AA   \\
  \hline
  \hline
  J$_{bb}(1)$ via O4, O10 & \\
    d(Mn1-Mn1) = 3.170(5)~\AA  &  \\
    Mn1-O4-Mn1 = 92.0(3)$^\circ$ & Mn1-O10-Mn1 = 99.0(3)$^\circ$ \\
  \hline
  J$_{bb}(2)$ via O1, O11  & \\
   d(Mn1-Mn1) = 3.106(5)~\AA & \\
   Mn1-O1-Mn1 = 88.2(3)$^\circ$ & Mn1-O11-Mn1 = 99.2(3)$^\circ$ \\
  \hline
  J$_{bv}$ via O5, O10  & \\
    d(Mn1-Mn2) = 3.376(4)~\AA & \\
    Mn1-O5-Mn2 = 91.5(2)$^\circ$ & Mn1-O10-Mn2 = 107.9(3)$^\circ$ \\
 \hline
 \end{tabular}
\caption{Bond lengths and angles around Mn ions in Rb$_2$Mn$_3$(MoO$_4$)$_3$(OH)$_2$~\cite{pellizzeri2019alkali}.}
\label{Tab:bond}
\end{table}
\endgroup

Temperature-dependent magnetization measurements were performed using a Quantum Design Magnetic Property Measurement System. A 5.3~mg single crystal was used for the measurements. The single crystal was affixed to a quartz rod using GE varnish and the magnetic properties perpendicular to flat surface (the normal direction is along the $c$-axis) of the single crystal was performed by sandwiching the single crystal between two quartz rods. Temperature dependent magnetic susceptibility was performed from 2 to 300~K in the applied magnetic field of 10~kOe.  Heat-capacity measurements were performed using Quantum Design Physical Property Measurement System. Single crystal neutron diffraction experiments were carried out at the CORELLI spectrometer~\cite{ye2018implementation} at the Spallation Neutron Source (SNS).  CORELLI is a quasi-Laue time-of-flight instrument with the incident neutron wavelength band between 0.65~\AA~and~2.9~\AA. It has a large 2D detector array, with a $-20^{\circ}$ to $+150^{\circ}$ in-plane coverage and $\pm28^{\circ}$ out-of-plane coverage. The sample was mounted with the (0,~k,~l) plane horizontal and the vertical rotation axis is along the $a$-axis. Experiments were conducted by rotating the sample for $\sim 120^\circ$ with a 2$^\circ$ step at the base temperature of 1.5~K. Totally 515 Bragg peaks collected at 1.5~K were used for the refinement, including 331 main reflections and 184 satellites. The Mantid package was used for data reduction including Lorentz and spectrum corrections~\cite{michels2016expanding}. The integrated Bragg intensities were obtained from integration in the 3D reciprocal space and were corrected for background. Temperature dependence studies were performed at a couple of selected angles optimized for signals from selected Bragg peaks of interest between 1.50~K and 5.50~K.

Possible magnetic structures were first investigated by the representation analysis method using the SARAh program~\cite{wills2000new}. This methodology was pioneered by Bertaut~\cite{bertaut1968representation} and Izyumov\cite{Izyumov1991neutron}, where the magnetic structure models are generated by summing over independent basis modes that transform according to active irreducible representations (irreps) of the space group of the paramagnetic phase~\cite{rodriguez2012symmetry, rodriguez2019magnetic}. Chemical and magnetic structural refinements were then carried out with the FullProf Suite program~\cite{rodriguez1993recent}.  To solidify the refinement results, the magnetic structures were further explored with the magnetic space and superspace group (a.k.a. Shubnikov and Shubnikov superspace groups) approaches, which give a phenomenological description of magnetic structures. Providing the magnetic moment of one magnetic atom, the symmetry operations of the magnetic group will generate magnetic moments of all other atoms in the same magnetic orbit. At the same time, each spin component is subject to site-symmetry constraints if the magnetic atom is not in a general position. For the commensurate magnetic structure, maximal magnetic space groups were found with the MAXMAGN program~\cite{perez2015symmetry}, and the generated magnetic structure models were refined with the FullProf Suite program~\cite{rodriguez1993recent}. For the incommensurate magnetic phase, the magnetization density distribution breaks the translation symmetry of the nuclear structure in the 3D space. However, the modulated magnetic structure can be mapped into an artificial space with a higher dimension using the superspace approach~\cite{janner1980symmetry, deWolff:a19729, perez2012magnetic}, where the translational symmetry is recovered by introducing new variables, i.e., the internal phase shifts that represent translations along the additional superspace coordinate axes. Experimental data were analyzed using the magnetic superspace group approach implemented in the software JANA2006~\cite{perez2012magnetic, petvrivcek2014crystallographic} and the on-line software ISODISTORT~\cite{stokes2006isodisplace}. The magnetic superspace group reported from the JANA2006 refinement were updated to the standard setting using the on-line software FINDSSG~\cite{stokes2011generation, van2013equivalence}.

\section{magnetic susceptibility and heat capacity} \label{sec:chi}

\begin{figure}[tb]
	\centering
		\includegraphics[width=0.48\textwidth]{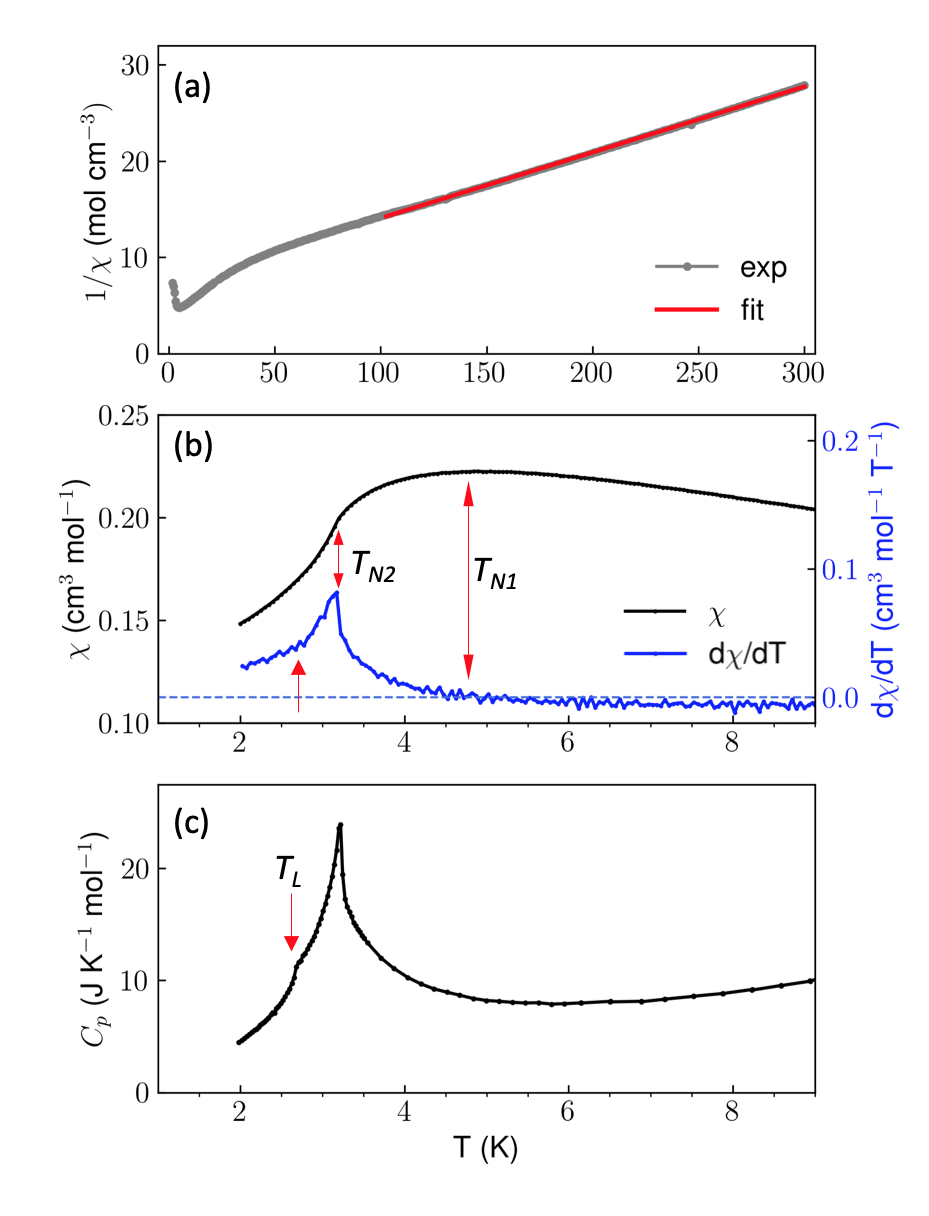} 
		\caption{ \label{Fig:chi} (Color online) (a) Temperature dependence of magnetic inverse susceptibility data and the best fit to the Curie-Weiss model in the high temperature region. It shows an effective moment of 5.9(1)~$\mu_B$ per Mn ion and a Weiss temperature of -105.9(4)~K.   (b) Low temperature portion of the magnetic susceptibilities. Two obvious anomalies were observed near 4.7~K and 3.2~K, respectively. Magnetic susceptibility data were collected in an applied magnetic field of 10~kOe. (c) Temperature dependence of zero-field heat capacity shows a $\lambda$-shape anomaly at 3.2~K and a kink around 2.6~K.}
\end{figure}

Magnetization measurements were performed on single crystals with the long-edge aligned parallel to the applied magnetic field. Figure~\ref{Fig:chi}(a) shows the temperature dependence of inverse magnetic susceptibility $\chi^{-1}$. There is a weak anisotropy in magnetic susceptibility between the field parallel and perpendicular to the long edge (data not shown). The high temperature portion of the inverse susceptibility was fitted with the Curie-Weiss model, $\chi = C/(T~-~\Theta_{CW})$, where $\Theta_{CW}$ is the Weiss constant. From the best fit, an effective moment of $5.9(1)~\mu_B$ per Mn ion and a Weiss temperature of -105.9(4)~K are determined. The effective moment is consistent with the expected spin-only value $5.91~\mu_B$ per Mn for high spin d$^{5}$ Mn$^{2+}$ ions ($S = 5/2$). The large and negative Weiss temperature indicates strong antiferromagnetic interactions. There are clear anomalies at low temperatures. To better view these, Fig.~\ref{Fig:chi}(b) shows the magnetic susceptibility of the low temperature region, where $\chi$ shows a downturn on cooling below 4.7~K, suggesting the onset of a long-range antiferromagnetic magnetic order below T$_{N1} \approx$~4.7~K. An anomalous slope change also exists around T$_{N2} \approx$~3.2~K, suggesting a second magnetic phase transition. These two anomalies can be easily seen in $d\chi/dT$, where $d\chi/dT$ changes sign and shows a peak at T$_{N1}$ and T$_{N2}$, respectively. The 3.2-K anomaly is obvious from the heat capacity measurement too, where it gives rise to a $\lambda$-shape peak, as shown in Fig.~\ref{Fig:chi}(c).  The heat capacity shows an additional kink at 2.6~K, which corresponds to a slope change in $d\chi/dT$ and indicates some sudden changes in the magnetic structure. Note that $\Theta_{CW}$ is significantly higher than T$_{N}$'s, suggesting strongly frustrated interactions with a frustration parameter $ f = |\Theta_{CW}|/T_{N1} \approx 24$.

\section{Magnetic Structures from Single crystal neutron diffraction}
\subsection{temperature dependence of magnetic wavevectors}  \label{subsec:op}

\begin{figure}[tb]
	\centering
		\includegraphics[width=0.45\textwidth]{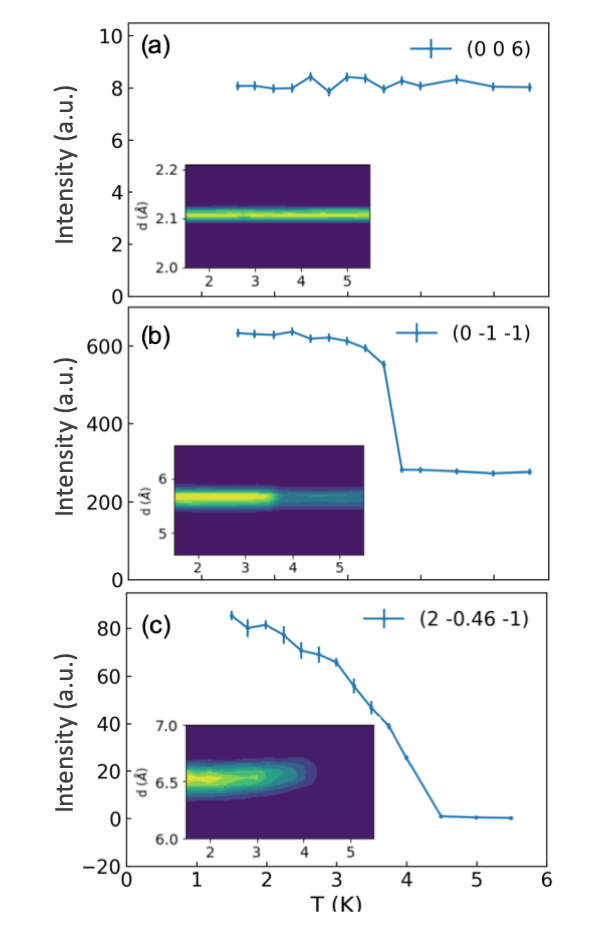} 
		\caption{ \label{Fig:OPs} (Color online) Temperature dependence of integrated peak intensities and the peak profiles of selected Bragg peaks. Lorentz and spectrum corrections were not performed. (a) Integer Bragg peak $(0,0,6)$ shows no obvious anomaly. (b) Integer Bragg peak $(0,-1,-1) $shows a large enhancement of intensity below 3.50~K, indicating a magnetic scattering contribution associated with a commensurate magnetic wavevector $(0,0,0)$. (c) Incommensurate Bragg peak near $(2,-0.46,-1)$ shows the onset of long-range incommensurate magnetic order below 4.20~K. Insets show the temperature dependence of the corresponding peak profiles. }
\end{figure}

To clarify the nature of the anomalies observed in the magnetic susceptibility and the heat capacity data, we performed single crystal neutron diffraction experiments between 1.5~K and 5.5~K with a fine temperature step size of 0.25-0.50~K. Figure~\ref{Fig:OPs} shows the representative 2D slices in the ($0,~k,~l$) plane at temperatures below T$_{N2}$, between T$_{N1}$ and T$_{N2}$, and above T$_{N2}$. It is obvious that features inside the circles show strong temperature dependence. Inside the small circles, the commensurate Bragg peak is clearly visible at 1.50~K, but significantly dimmed at 3.75~K and 5.50~K. Inside the big circles, there are incommensurate Bragg peaks with a modulation wavevector around ($0,~0.46,~0$) at 1.50~K and 3.75~K, which are absent at 5.5~K. The incommensurate Bragg peaks become weaker at locations with both a high $K$ and a high $L$ indices, suggesting their magnetic scattering origin. This is more obvious in the mesh scan dataset shown below in Fig.~\ref{Fig:mesh}(a). 

\begin{figure}[tb]
	\centering
		\includegraphics[width=0.45\textwidth]{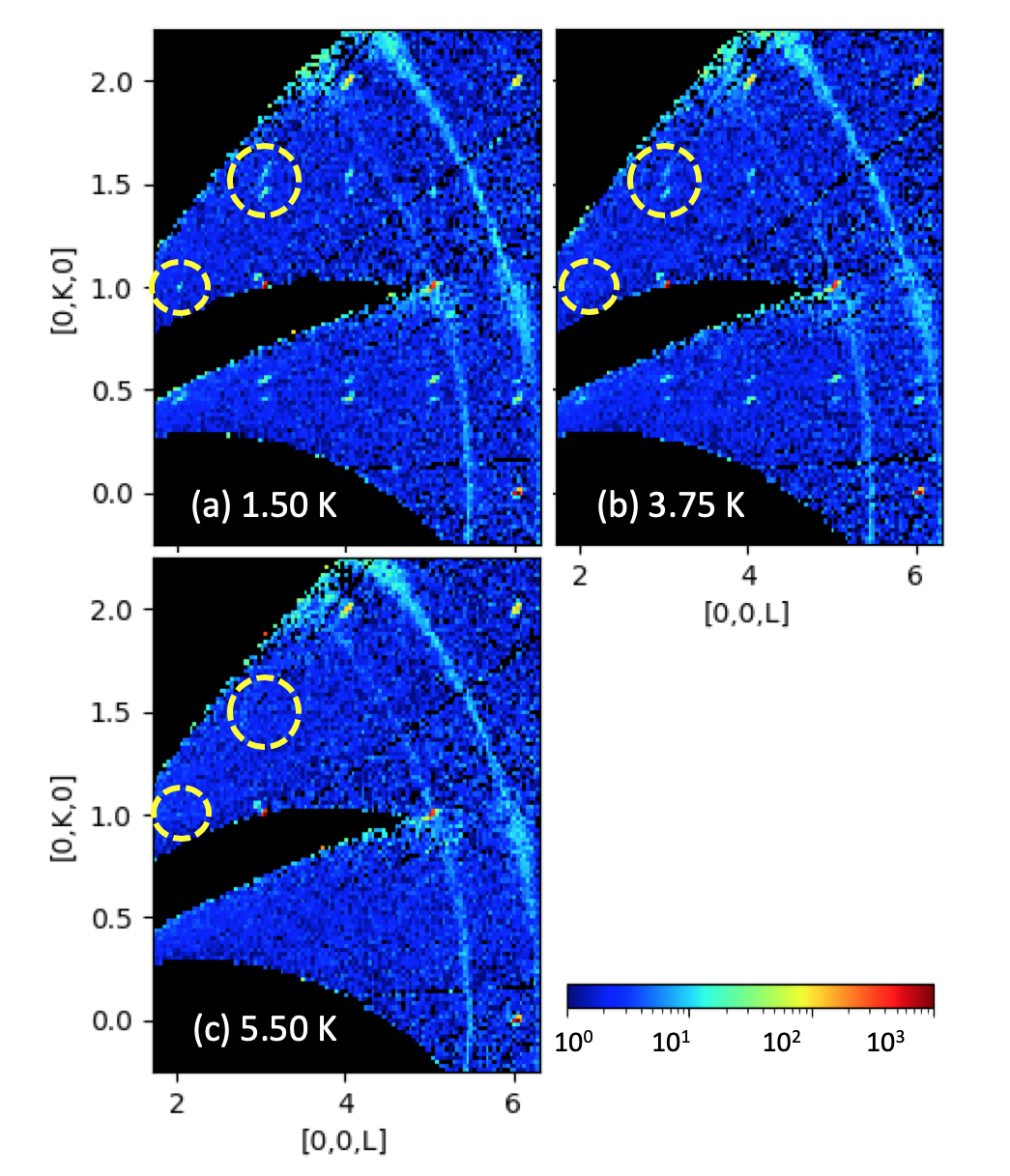} 
		\caption{ \label{Fig:OP_2D} (Color online) 2D slices in the ($0,~k,~l$) plane of the neutron diffraction data at selected temperatures, (a) 1.50~K, (b) 3.75~K and (c) 5.50~K, respectively. Inside circles are selected features with strong temperature dependence. Inside the big circles, there are additional incommensurate peaks at 1.50~K and 3.75~K, which is absent at 5.50~K. Inside the small circles, the commensurate Bragg peak is clearly visible at 1.50~K, but significantly dimmed at 3.75~K and 5.50~K. Also visible are the powder rings from scattering of the sample environment and sample mount. }
\end{figure}

\begin{figure}[tb]
	\centering
		\includegraphics[width=0.4\textwidth]{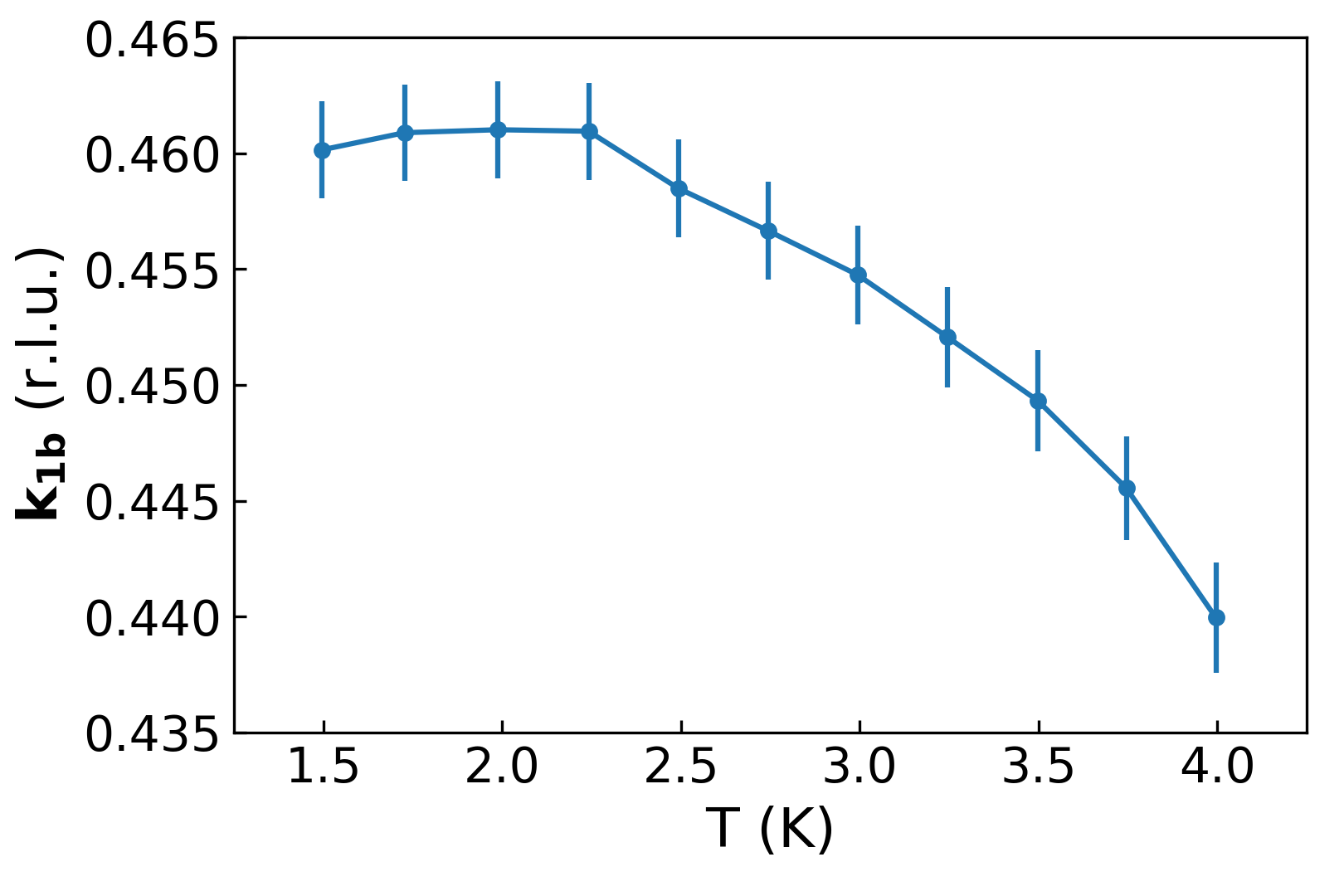} 
		\caption{ \label{Fig:icmkb} (Color online) Temperature dependence of the incommensurate wavevector ${\bf k}_{1}$ measured at the Bragg position (2, -${\bf k_{1b}}$, -1).}
\end{figure}

Figure~\ref{Fig:OPs} shows the temperature dependence of the integrated peak intensities and the peak profiles (inset) for selected peaks. The integrated peak intensities were achieved by peak profile fitting against the $d$-spacing value after converting the data from the detector space to the $d$-space. To minimize uncertainties, Lorentz and spectrum corrections were not performed; therefore, the intensities from different peaks are not directly comparable. The integer Bragg peak ($0,~0,~6$) shows no obvious anomaly on cooling, suggesting no structure transformation occurring at low temperatures. The low-$Q$ Bragg peak ($0,~-1,~-1$) shows a large enhancement of intensity below 3.50~K, indicating a magnetic scattering contribution associated with a commensurate wavevector ($0,~0,~0$). The incommensurate Bragg peak near ($2,~-0.46,~-1$) shows the onset of long-range magnetic order below 4.5~K. From the inset of Fig.~\ref{Fig:OPs}(c), a small but notable change in the $d$-spacing corresponds to the incommensurate peak. After transforming the data into the reciprocal space (Fig.~\ref{Fig:icmkb}), it is clearly seen that the incommensurate wavevector changes as a function of temperature and undergoes a lock-in transition below 2.3~K. Therefore, the anomalies observed in the heat capacity and the magnetic susceptibility data discussed in Sec.~\ref{sec:chi} correspond to various magnetic structure transitions found in the neutron diffraction data. Overall, the sample experiences a paramagnetism to incommensurate-antiferromagnetism phase transition at 4.5~K on cooling, and then forms a complex magnetic structure with two wavevectros from different orbits below 3.5~K. On further cooling, the incommensurate wavevector undergoes a lock-in transition below 2.3~K.

\subsection{incommensurate magnetically ordered component with \texorpdfstring{${\bf k}_{1}~\approx~(0,~0.46,~0)$}{}} \label{subsec:inc}

\begin{figure}[tb]
	\centering
		\includegraphics[width=0.48\textwidth]{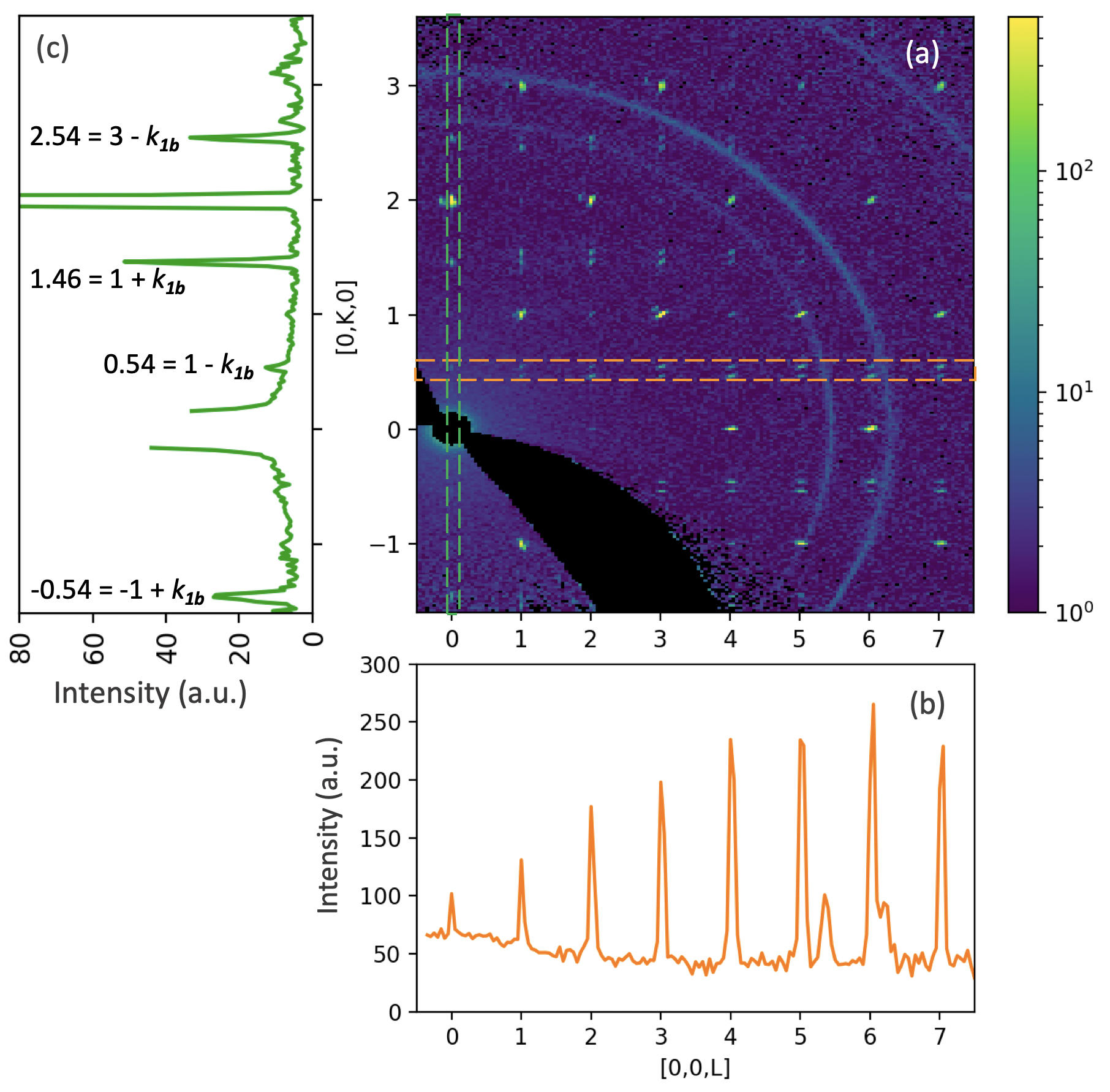} 
		\caption{ \label{Fig:mesh} (Color online) (a) The slice-cut in the ($0, k, l$) plane of the neutron diffraction data collected at 1.50~K. (b) A line cut along the $L$ direction, with a integration cross section of $K =[0.4, 0.6]$ (as shown by the dashed line in (a)) and $H = [-0.05, 0.05]$. (c)  The satellite peaks along $[0, K, 0]$ direction show a selection-rule-like $K$ dependence. The peaks associated with odd integer $K$s are pronounced, and the peaks associated with even integer $K$s are barely observable.}
\end{figure}

A mesh scan by rotating the sample through 120$^\circ$ was conducted at the base temperature of 1.50~K to cover both [$0~0~L$] and [$0~K~0$] axes. Figure~\ref{Fig:mesh}(a) shows the slice-cut in the ($0,~k,~l$) plane. Interestingly, the satellite peak intensity shows a non-monotonic dependence on $L$ for small $K$'s. As shown by the line-cut in Fig.~\ref{Fig:mesh}(b), the peak intensity increases first as $L$ increases before it eventually decreases. For an orthorhombic lattice, such a dependence suggests that the relevant ordered magnetic moment has a much larger component in the $ab$-plane than the component along the $c$-axis, because of the vectorial nature of the neutron-magnetic moment interaction~\cite{halpern1939magnetic}.  There are also a selection-rule-like feature for the satellite peaks along the $[0, K, 0]$ direction. As shown by the line-cut (Fig.~\ref{Fig:mesh}(c)),  they are very pronounced for odd integer $K$s, but barely observable for even integer $K$s. This feature has been used to distinguish magnetic structure models that give rise to comparable refinement results.

\begin{figure}[tb]
	\centering
		\includegraphics[width=0.48\textwidth]{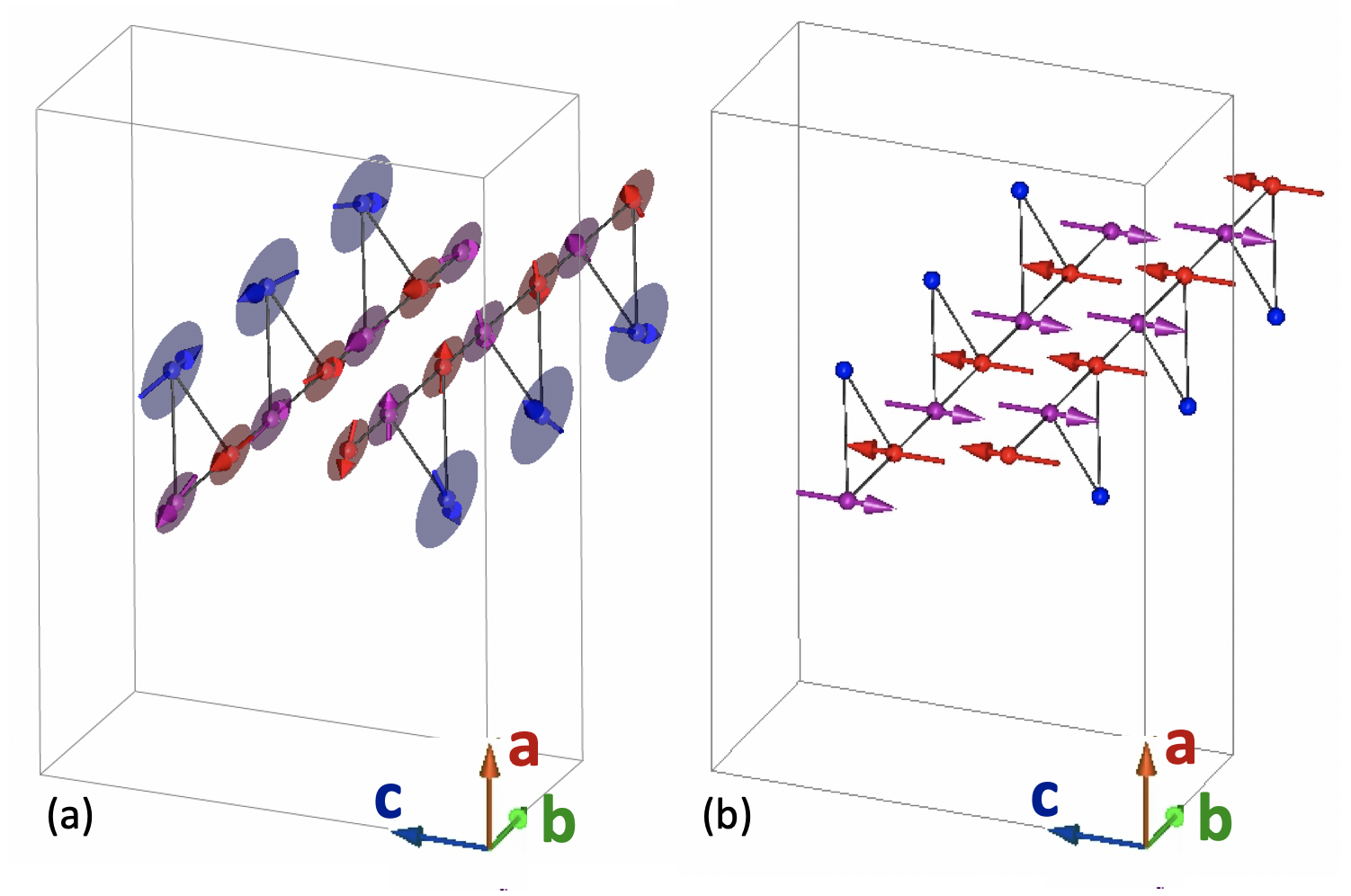} 
		\caption{ \label{Fig:simplemagstru} (Color online) The simplified magnetic structure describing the major ordered components, where the incommensurate component and commensurate component are mutually normal to each other. The results are obtained from the irrep modes via Fullprof. (a) The incommensurate component consists of cycloidally ordered magnetic moments lying in the $ab$-plane from both Mn1 and Mn2 sites. (b) The commensurate component consists of collinear spins on Mn1 sites with antiferromagnetically coupled moments along $b$-axis between adjacent Mn1 ions. The purple, red and blue colors label the Mn1a, Mn1b and Mn2 sites, respectively.  The moment sizes on Mn1 and Mn2 sites are not drawn to scale for the purpose of clarity. }
\end{figure}

Incommensurate magnetic structure models were investigated using the representation analysis with the program SARAh~\cite{wills2000new} to determine the symmetry-allowed magnetic structures that can result from a second-order magnetic phase transition given the crystal structure and the propagation vector. As mentioned above, there are two crystallographically different Mn sites, Mn1 at a $8d$ Wyckoff position and Mn2 at a $4c$ Wyckoff position. The little group $G_{\bf{k1}}$ of the propagation vector ${\bf k}_{1} = (0,~0.46,~0)$ only contains 4 out of the 8 symmetry operators of the space group $G_0$. Under the little group $G_{k1}$, Mn1 sites are separated into 2 orbits, i.e., Mn1a and Mn1b sites, alternately sitting on the spin chain, as illustrated in Fig.~\ref{Fig:simplemagstru}a by the purple and red colors, respectively. The Mn1a and Mn1b Wyckoff orbits are connected by the space inversion symmetry, and the splitting is because the symmetry mode analysis adapted by the Sarah program does not take into account the space inversion symmetry operation that transforms $\bf{k}$ into -$\bf{k}$~\cite{perez2012magnetic, harris2007landau}.

\begingroup
\renewcommand{\arraystretch}{0.5}
\begin{table}[tb]
\begin{tabular}{ccc|cccccc}
\hline\hline
      &  BV  &  Atom & \multicolumn{6}{c}{BV components}\\
      &      &             &$m_{\|a}$ & $m_{\|b}$ & $m_{\|c}$ &$im_{\|a}$ & $im_{\|b}$ & $im_{\|c}$ \\
\hline
Mn1a         & $\bf \psi_{1}$ &      1 &      1 &      0 &      0 &      0 &      0 &      0  \\
             &              &      2 &   .125 &      0 &      0 &   .992 &      0 &      0  \\
             &              &      3 &   .125 &      0 &      0 &   .992 &      0 &      0  \\
             &              &      4 &      1 &      0 &      0 &      0 &      0 &      0  \\
             & $\bf \psi_{2}$ &      1 &      0 &      1 &      0 &      0 &      0 &      0  \\
             &              &      2 &      0 &  -.125 &      0 &      0 &  -.992 &      0  \\
             &              &      3 &      0 &  -.125 &      0 &      0 &  -.992 &      0  \\
             &              &      4 &      0 &      1 &      0 &      0 &      0 &      0  \\
             & $\bf \psi_{3}$ &      1 &      0 &      0 &      1 &      0 &      0 &      0  \\
             &              &      2 &      0 &      0 &   .125 &      0 &      0 &   .992  \\
             &              &      3 &      0 &      0 &  -.125 &      0 &      0 &  -.992  \\
             &              &      4 &      0 &      0 &     -1 &      0 &      0 &      0  \\
\hline
Mn1b         & $\bf \psi_{1}$ &      1 &      1 &      0 &      0 &      0 &      0 &      0  \\
             &              &      2 &  -.368 &      0 &      0 &   -.93 &      0 &      0  \\
             &              &      3 &  -.368 &      0 &      0 &   -.93 &      0 &      0  \\
             &              &      4 &      1 &      0 &      0 &      0 &      0 &      0  \\
             & $\bf \psi_{2}$ &      1 &      0 &      1 &      0 &      0 &      0 &      0  \\
             &              &      2 &      0 &   .368 &      0 &      0 &    .93 &      0  \\
             &              &      3 &      0 &   .368 &      0 &      0 &    .93 &      0  \\
             &              &      4 &      0 &      1 &      0 &      0 &      0 &      0  \\
             & $\bf \psi_{3}$ &      1 &      0 &      0 &      1 &      0 &      0 &      0  \\
             &              &      2 &      0 &      0 &  -.368 &      0 &      0 &   -.93  \\
             &              &      3 &      0 &      0 &   .368 &      0 &      0 &    .93  \\
             &              &      4 &      0 &      0 &     -1 &      0 &      0 &      0  \\

\hline
Mn2          & $\bf \psi_{1}$ &      1 &      1 &      0 &      0 &      0 &      0 &      0  \\
             &              &      2 &  -.368 &      0 &      0 &   -.93 &      0 &      0  \\
             &              &      3 &  -.368 &      0 &      0 &   -.93 &      0 &      0  \\
             &              &      4 &      1 &      0 &      0 &      0 &      0 &      0  \\
             & $\bf \psi_{2}$ &      1 &      0 &      1 &      0 &      0 &      0 &      0  \\
             &              &      2 &      0 &   .368 &      0 &      0 &    .93 &      0  \\
             &              &      3 &      0 &   .368 &      0 &      0 &    .93 &      0  \\
             &              &      4 &      0 &      1 &      0 &      0 &      0 &      0  \\
             & $\bf \psi_{3}$ &      1 &      0 &      0 &      1 &      0 &      0 &      0  \\
             &              &      2 &      0 &      0 &  -.368 &      0 &      0 &   -.93  \\
             &              &      3 &      0 &      0 &   .368 &      0 &      0 &    .93  \\
             &              &      4 &      0 &      0 &     -1 &      0 &      0 &      0  \\

\hline
\end{tabular}
\caption{Basis vectors for the space group $Pnma$ (\#62) with ${\bf k}_{1} = (0,~0.46,~0)$ associated with the IR $\Gamma_{3}$ for Mn1a (0.55, 0.49, 0.32), Mn1b (0.95, 0.51, 0.82) and Mn2 (0.40, 0.75, 0.25) sites. Note that Mn1a and Mn1b are connected with the space group symmetry operation of  $\{-1 \vert 0\} \times \{ m_{001} \vert {\frac{1}{2}}~0~\frac{1}{2}\}$, i.e., ($-x-\frac{1}{2}, -y, z-\frac{1}{2}$), with an additional translation $\bf{t}$~=~(2,1,1). Alternatively, they are connected by $\{ 2_{001} \vert {\frac{1}{2}}~0~\frac{1}{2}\}$, i.e., ($-x+\frac{1}{2}, -y, z+\frac{1}{2}$), with an additional translation $\bf{t}$~=~(1,1,0). }
\label{Tab:Mnbvs}
\end{table}
\endgroup

There are four one-dimensional irreps associated with the  little group $G_{\bf{k1}}$. The decompositions of the magnetic representation for the three Mn sites (Mn1a, Mn1b and Mn2) have the same form,  $\Gamma_{Mag}=3\Gamma_{1}^{1}+3\Gamma_{2}^{1}+3\Gamma_{3}^{1}+3\Gamma_{4}^{1}$. The labeling of the IRs follows the scheme used by Kovalev~\cite{kovalev1993representations}. For all these IRs, there is no restriction on the magnetic moment directions on any Mn atoms. Simulated annealing was used to prescreen potential models based on individual single IR. Both the spin density wave (SDW) model and the cycloid model based on $\Gamma_{3}$ fit the experimental data significantly better than other models. Table~\ref{Tab:Mnbvs} lists the basis vectors $\bf \psi_{i}$ associated with the IR $\Gamma_{3}$ for the three Mn sites. Refinement shows that the cycloid model is slightly better than the SDW model.  

Figure~\ref{Fig:simplemagstru}a shows the incommensurate magnetic structure generated from the best model. As mentioned above, the operations that interchange $\bf{k}$ and -$\bf{k}$ was overlooked in our representation analysis,  therefore it allows the independent magnetization modulations for Mn1a and Mn1b sites, a pair of atoms related by the space inversion operation~\cite{harris2007landau, perez2012magnetic}. During the refinement, the amplitudes of the Mn1a and Mn1b sites were constrained to be equal since they are on chemically equivalent sites, but there was no constraint on their relative orientations. As mentioned above, the vectorial magnetization distribution are expressed as a linear combination of the mutually orthogonal basis vectors. Therefore, there are multiple combinations that can constraint the amplitudes of the Mn1a and Mn1b sites to be equal, such as  (1) $m^{x}_{Mn1a} = m^{x}_{ Mn1b}$, $m^{y}_{Mn1a} = m^{y}_{Mn1b}$ and $m^{z}_{Mn1a} = m^{z}_{Mn1b}$, or (2)  $m^{x}_{Mn1a} = m^{x}_{ Mn1b}$, $m^{y}_{Mn1a} = -m^{y}_{Mn1b}$ and $m^{z}_{Mn1a} = m^{z}_{Mn1b}$, and so on. They correspond to the different magnetic structures. Without considering the symmetry operation connecting these two sites, we need to test one by one against the experimental data. The refinement shows essentially no improvement by introducing non-zero moment along the $c$-axis on any Mn sites. This agrees with the $L$-dependence of the incommensurate peak intensities discussed above. Therefore, the magnetic moment directions are constrained in the $ab$-plane. The best model shows that $m^{x}_{Mn1a}/m^{x}_{ Mn1b}$ and $m^{y}_{Mn1a}/m^{y}_{ Mn1b}$ have opposite signs.  The cycloids described by both Mn moments have an elliptical envelope, which implies that the moment amplitudes are oscillating across the $ab$ plane. By fixing the phase of the Mn1a atom at (0.55, 0.48, 0.32) to be $\Phi_{Mn1a} = 0$, the optimized phase of the Mn1b atom at (0.45, 0.52, 0.82) is $\Phi_{Mn1b} = 0.243(8) \times 2\pi$ and the phase of the Mn2 atom at (0.40, 0.75, 0.25)  is $\Phi_{Mn2} = 0.46(2) \times 2\pi$. Note that neutron diffraction data collected at 1.5~ K suggestions a slight shift of the Mn1 positions with respect to the room temperature x-ray diffraction result.

\begingroup
\renewcommand{\arraystretch}{0.64}
\begin{table}[tb]
\centering
\begin{tabular}{|C{1.0cm}|C{1.4cm}|C{1.4cm}|C{1.4cm}|C{1.4cm}|C{1.4cm}|}
  \hline
  & \multicolumn{3}{C{4.0cm}|}{${\bf k}_{1}~\approx~(0,~0.46,~0)$} & \multicolumn{2}{C{2.8cm}|}{${\bf k}_{2}~\approx~(0,~0,~0)$} \\
  \hline
  & $m_a$ & $m_b$ &phase& $m_c$ & $m_c$$^*$ \\
  \hline
  Mn1a &0.99(9)&1.99(7)& 0            & 3.10(16) & 2.93(7)\\
  \hline
  Mn1b &0.99(9)&-1.99(7)&0.243(8) &-& -\\
  \hline
  Mn2 &2.55(8)&4.59(5) &0.46(2)     & 0 & 0 \\
  \hline
 \end{tabular}
\caption{Summary of the ordered moment components (in  $\mu_B$ per ion) and phase (in unit of $2*\pi$ at 1.5~K from the simplified model (the irrep mode approach) associated with ${\bf k}_{1}$ and ${\bf k}_{2}$, respectively.  $m_c$$^*$ was obtained by fitting the difference dataset between $1.8 \pm 0.3$~K and  and $4.8 \pm 0.6$~K.}
\label{Tab:om}
\end{table}
\endgroup

We have further used the magnetic superspace group approach to identify the potential models of the incommensurate magnetic structure, using the software JANA2006~\cite{perez2012magnetic, petvrivcek2014crystallographic}, and the software ISODISTORT~\cite{stokes2006isodisplace}.  Recent development of the superspace algorithm have shown that the superspace symmetry will generally introduce either stricter or equivalent restrictions on the magnetic structures than the representation method alone~\cite{perez2012magnetic}. In the superspace formalism~\cite{deWolff:a19729}, the modulated magnetic moment of an atom located at the position $r_{\nu}$ in the unit cell of the basic structure can be expressed as a Fourier series:
\begin{equation}
\begin{aligned}
\label{eq:ssg}
\vec{M}_{\nu}(\vec{k}~\cdot~\vec{r}_{\nu}) = \vec{M}_{\nu0}  + \sum_{m}  [\vec{M}_{\nu, ms} sin(2 \pi m \vec{k}~\cdot~\vec{r}_{\nu}) \\ +  \vec{M}_{\nu, mc} cos(2 \pi m \vec{k}~\cdot~\vec{r}_{\nu}) ], 
\end{aligned}
\end{equation}
where  $\vec{M}_{\nu0}$ is the absolute term, and $\vec{M}_{\nu, ms}$ and $\vec{M}_{\nu, mc}$ are the amplitudes of the sine and cosine terms, respectively. The first term ($m=0$) will contribute to scattering intensity to main reflections and the harmonic terms ($m=1, 2, ...$) will give rise to magnetic satellite peaks. We only observed first harmonics peaks for the single-$\bf{k}$ incommensurate phase,  therefore it is sufficient to only consider $m = 1$ terms. The Fourier components can be decomposed into three components along the principal crystallographic directions, and each spin component will be subjected to site-symmetry constraints if the magnetic atom is not in a general position.  Jana2006 reports four primary superspace groups for the extend little group $G_{\bf{k, -k}}$ ($\bf{k}~= $~(0, 0.46, 0)), each of which corresponds to a two-dimensional representation. They are $Pnma.1'(0b0)000s$, $Pnma.1'(0b0)s0ss$, $Pnma.1'(0b0)s00s$, and $Pnma.1'(0b0)00ss$, respectively.   The superspace groups $Pnma.1'(0b0)s0ss$ and $Pnma.1'(0b0)00ss$ give significantly better refinement results than the other two. $Pnma.1'(0b0)s0ss$ gives the best refinement result and reproduces the reflection condition for satellite peaks along $[0, K, 0]$, as observed experimentally (Fig.~\ref{Fig:mesh}(c)). The symmetry constraints of $Pnma.1'(0b0)s0ss$ on the spin components as well as the final refined results are shown in ~\ref{Tab:icmSSG}.

\begin{figure}[tb]
	\centering
		\includegraphics[width=0.96\textwidth]{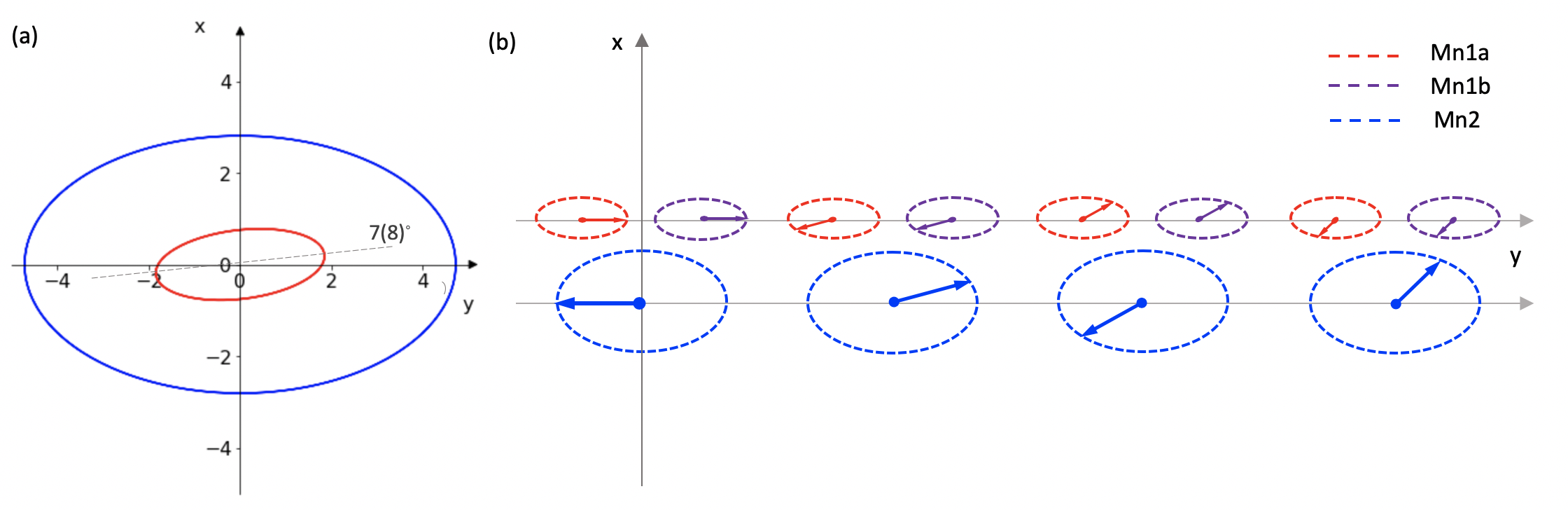} 
		\caption{ \label{Fig:MnSSG} (Color online)  (a) The $ab$-plane moment trajectories of the incommensurate component of Mn1 and Mn2 sites, as determined from the best fit using the magnetic superspace approach via Jana2006, describe elliptical rotations. (b) The approximate incommensurately ordered magnetic moments of the Mn1 (red and purple) and Mn2 (blue) sites within one spin chain. Within the experimental uncertainty, the largest moments for both Mn1 and Mn2 sites are along the $b$-axis.}
\end{figure}

The refinement show that $m_{cz}$ and $m_{sz}$ are vanishingly small for both Mn sites, validating the simplified model only considering ordered moments in the $ab$ plane. From the amplitudes of the sine and cosine components, it is easy to find that the spin modulation along the $a$ and $b$ axes are out-of-phased by $\sim 75^\circ$ for Mn1 sites, and $90^\circ$  and Mn2 sites, respectively. This corresponds to cycloid spin structures rather than SDW structures for both sites, which strengthens the previous results based on the irrep mode analysis. The magnetic moments of the Mn1 and Mn2 sites are elliptically rotating in the $ab$-plane, as shown in Fig.\ref{Fig:MnSSG}.  By ignoring the small order moment along the $c$-direction, the moment trajectories of Mn1 sites and M2 sites are approximately elliptical cycloids within the $ab$-plane. For Mn1 sites, the largest moment is 1.88(11) $\mu_B$/Mn and tilted by an angle of $6.8(7.7)^\circ$ away from the $b$-axis;  and for Mn2 sites,  the largest moment is 4.72(16) $\mu_B$/Mn and is along $b$-axis. These values agree with those from the simplified model using the irrep modes within uncertainties. Figure~\ref{Fig:MnSSG}b shows that within each spin triangle, the moment directions of Mn1a and Mn1b are parallel to each other in the $ab$-plane (by the symmetry constraint), and the moment of Mn2 is antiparallel to those of Mn1 (by refinement).

\begingroup
\renewcommand{\arraystretch}{0.64}
\begin{table*}[tb]
\begin{tabular}{|C{1.8cm}|C{1.64cm}|C{1.64cm}|C{1.64cm}|C{1.64cm}|C{1.64cm}|C{1.64cm}|C{1.64cm}|C{1.64cm}|}
\hline
              \multicolumn{9}{|c|}{$\bf{k_2}$ = (0, 0, 0), m = 0 term}                                                   \\
\hline
Atom             & \multicolumn{4}{c|}{Mn1}                                                          & \multicolumn{4}{c|}{Mn2}                   \\
\hline
                      &   x              & y               & z                & length                      &  x              & y               & z              & length \\
\hline
constraints     &m$_{x0}$   & m$_{y0}$  & m$_{z0}$  &                                & m$_{x0}$ & m$_{y0}$  & m$_{z0}$ &           \\
\hline
value             &  -0.49(11)   & 1.08(8) & -3.53(9)  & 3.73(16)                   & 0               & -0.16(15)   &0               &0.16(15) \\
\hline
\hline
               \multicolumn{9}{|c|}{$\bf{k_1}$ = (0, 0.46, 0), m = 1, first harmonic term}  \\
\hline             
Atom            &  \multicolumn{4}{c|}{Mn1, cosine component}                       &  \multicolumn{4}{c|}{Mn2, cosine component}  \\
\hline
                      &     x                   & y                  & z                   & length      &          x            &     y             &   z              & length         \\
\hline
constraints     &  m$_{cx}$      & m$_{cy}$     & m$_{cz}$        &                &            0         & m$_{cy}$    & 0   &           \\
\hline
value              &  -0.53(18)         &  1.02(12)     & 0.024(20)       & 1.15(29)  &        0           &      -4.72(16)            &    0 &  4.72(16)         \\
\hline
 Atom           &\multicolumn{4}{c|}{Mn1, sine component}                             &\multicolumn{4}{c|}{Mn2, sine component}     \\
\hline
                      &   x                     &     y              &          z            & length    &          x          &           y        & z                  &  length\\
\hline
constraints    &       m$_{sx} $   &  m$_{sy}$    &      m$_{sz}$   &               &         m$_{sx}$   &  0    & m$_{sz}$                 &            \\
\hline
value             &      -0.57(21)       & -1.54(11)     &    0.04(23)    &1.64(33)  &         2.81(17)       &  0      & 0.02(54)            & 2.81(56)         \\
\hline

\end{tabular}
\caption{System constraints on the spin components and the refinement results (in $\mu_{B}$ per atom) from Jana2006 of the magnetic Mn atoms from the Shubnikov superspace group $Pn'ma(0b0)00s$. The model was refined against the dataset collected at 1.5~K, with 515 Bragg peaks, including 331 main reflections and 184 satellites. For the parent magnetic superspace group  $Pnma.1'(0b0)s0ss$,  $\vec{M}_{0}$ (m = 0 components)  for both sites is restricted to be zero, while the symmetry constrains on sine and cosine Fourier components are same as those of $Pn'ma(0b0)00s$.  }
\label{Tab:icmSSG}
\end{table*}
\endgroup

In addition to the magnetic superspace group (or its equivalent magnetic superspace group) listed by Jana2006, ISODISTORT lists a second isotropy subgroup with a lower symmetry for each of the four two-dimensional representations. The space inversion symmetry operator is not included in these additional superspace groups, which thus require two instances of the irrep modes from the same irrep and allow more complex magnetic structures. However, the higher-symmetry $Pnma.1'(0b0)s0ss$ superspace group is sufficient to describe the experimental results and further decreasing the symmetry is not guaranteed. Using the on-line software FINDSSG~\cite{stokes2011generation, van2013equivalence}, $Pnma.1'(0b0)s0ss$ was converted to the standard setting, $Pbnm.1'(00g)s00s$ (62.1.9.4.m442.2), which has a point group symmetry of $mmm1'$.

\subsection{commensurate magnetically ordered component with \texorpdfstring{${\bf k}_{2}~=~(0,~0,~0)$}{}} 
\label{subsec:comm}

\begingroup
\renewcommand{\arraystretch}{0.5}
\begin{table}[tb]
\begin{tabular}{ccc|cccccc}
\hline
\hline
      &  BV  &  Atom & \multicolumn{6}{c}{BV components}\\
      &      &             &$m_{\|a}$ & $m_{\|b}$ & $m_{\|c}$ &$im_{\|a}$ & $im_{\|b}$ & $im_{\|c}$ \\
\hline
Mn1  & $\bf \psi_{1}$ &      1 &      1 &      0 &      0 &      0 &      0 &      0  \\
             &              &      2 &      1 &      0 &      0 &      0 &      0 &      0  \\
             &              &      3 &      1 &      0 &      0 &      0 &      0 &      0  \\
             &              &      4 &      1 &      0 &      0 &      0 &      0 &      0  \\
             &              &      5 &     -1 &      0 &      0 &      0 &      0 &      0  \\
             &              &      6 &     -1 &      0 &      0 &      0 &      0 &      0  \\
             &              &      7 &     -1 &      0 &      0 &      0 &      0 &      0  \\
             &              &      8 &     -1 &      0 &      0 &      0 &      0 &      0  \\
             & $\bf \psi_{2}$ &      1 &      0 &      1 &      0 &      0 &      0 &      0  \\
             &              &      2 &      0 &     -1 &      0 &      0 &      0 &      0  \\
             &              &      3 &      0 &     -1 &      0 &      0 &      0 &      0  \\
             &              &      4 &      0 &      1 &      0 &      0 &      0 &      0  \\
             &              &      5 &      0 &     -1 &      0 &      0 &      0 &      0  \\
             &              &      6 &      0 &      1 &      0 &      0 &      0 &      0  \\
             &              &      7 &      0 &      1 &      0 &      0 &      0 &      0  \\
             &              &      8 &      0 &     -1 &      0 &      0 &      0 &      0  \\
             & $\bf \psi_{3}$ &      1 &      0 &      0 &      1 &      0 &      0 &      0  \\
             &              &      2 &      0 &      0 &     -1 &      0 &      0 &      0  \\
             &              &      3 &      0 &      0 &      1 &      0 &      0 &      0  \\
             &              &      4 &      0 &      0 &     -1 &      0 &      0 &      0  \\
             &              &      5 &      0 &      0 &     -1 &      0 &      0 &      0  \\
             &              &      6 &      0 &      0 &      1 &      0 &      0 &      0  \\
             &              &      7 &      0 &      0 &     -1 &      0 &      0 &      0  \\
             &              &      8 &      0 &      0 &      1 &      0 &      0 &      0  \\
\hline
Mn2 & $\bf \psi_{1}$ &      1 &      0 &      2 &      0 &      0 &      0 &      0  \\
             &              &      2 &      0 &     -2 &      0 &      0 &      0 &      0  \\
             &              &      3 &      0 &     -2 &      0 &      0 &      0 &      0  \\
             &              &      4 &      0 &      2 &      0 &      0 &      0 &      0  \\

\hline
\end{tabular}
\caption{Basis vectors for the space group $Pnma$ with ${\bf k}_{2}=(0,~ 0,~ 0)$ associated with the IR $\Gamma_{4}$ for the Mn1 and Mn2 sites.}
\label{Tab:Comm}
\end{table}
\endgroup

By comparing the scattering pattern below and above $T_{N2}$, we have found that when $K$ is an even integer, the magnetic scattering contribution to the Bragg peaks are vanishingly small, if any. Based on this observation, we can expect an antiferromagnetic magnetic structure associated with ${\bf k}_{2}~=~(0,~0,~0)$, which agrees with the magnetic susceptibility data. 

From the representation analysis, there are eight one-dimensional irreps associated with the $Pnma$  space group (\#62) and the little group $G_{\bf{k2}}$ of the propagation vector ${\bf k}_{2}~=~(0,~0,~0)$. The decomposition of the magnetic representation for the Mn$1$ site is $\Gamma_{Mag}=3\Gamma_{1}^{1}+3\Gamma_{2}^{1}+3\Gamma_{3}^{1}+3\Gamma_{4}^{1}+3\Gamma_{5}^{1}+3\Gamma_{6}^{1}+3\Gamma_{7}^{1}+3\Gamma_{8}^{1}$. There is no restriction on the magnetic moment direction for each possible IR associated with the Mn1 site. The decomposition of the magnetic representation for the Mn2 site is $\Gamma_{Mag}=1\Gamma_{1}^{1}+2\Gamma_{2}^{1}+2\Gamma_{3}^{1}+1\Gamma_{4}^{1}+1\Gamma_{5}^{1}+2\Gamma_{6}^{1}+2\Gamma_{7}^{1}+1\Gamma_{8}^{1}$.  $\Gamma_{2}$, $\Gamma_{3}$, $\Gamma_{6}$ and $\Gamma_{7}$ only allow a non-zero magnetic moment in the $ac$-plane; and $\Gamma_{1}$, $\Gamma_{4}$, $\Gamma_{5}$ and $\Gamma_{8}$ only allow a non-zero magnetic moment along the $c$-axis.  From the magnetic symmetry approach, there are 8 possible maximal magnetic space groups. There is a one to one correspondence between the IR and the maximal magnetic space group for this case. From IR $\Gamma_{1}$ to $\Gamma_{8}$, they correspond to magnetic space groups $Pnma$(\#62.441), $Pn'm'a'$(\#62.449), $Pnm'a'$(\#62.447), $Pn'ma$(\#62.443), $Pn'ma'$(\#62.448), $Pnm'a$(\#62.444), $Pn'm'a$(\#62.446) and $Pnma'$(\#62.445), respectively.    

Various models based on each individual IR were fitted against the experimental data. We first refined both the nuclear structure and the commensurately ordered magnetic component simultaneously using the main reflection data collected at 1.5~K, which consists of both nuclear scattering and magnetic scattering contributions. For the incommensurate magnetic structure, the ordered magnetic moments mainly lie in the $ab$-plane with no abrupt change of the incommensurate peak intensities near $T_{N2}$. Therefore, it is reasonable to consider that the commensurate magnetization component has additionally ordered moments aligned along the $c$-axis. Refinement shows that only the magnetic structure described by the IR $\Gamma_{4}$ (magnetic space group $Pn'ma$(\#62.443)) agrees with the experimental data. Table~\ref{Tab:Comm} lists the basis vectors for the IR $\Gamma_{4}$, and Fig.~\ref{Fig:simplemagstru}b shows the refined magnetic structure model. We have further relaxed the model to allow the magnetic moments along all symmetry-allowed directions and used simulated annealing to search for potential models. However, no significant improvement has been found. To better separate the nuclear and the magnetic contributions to the Bragg peak intensities, we also refined the nuclear and magnetic structures separately using two datasets. The nuclear model was fitted against the data collected above $T_{N2}$ and the magnetic model was fitted against the difference between the data collected below and the above $T_{N2}$, respectively. The high-temperature and low-temperature datasets were collected at $4.8 \pm 0.6$~K and $1.8\pm0.3$~K, respectively. As expected, we obtained the same magnetic structure but slightly reduced ordered moment. The refined ordered moment values are listed in Tab.~\ref{Tab:om}. 

\subsection{two-\texorpdfstring{${\bf k}$}{} magnetic structure at low temperatures} 
\label{subsec:2k}

In Sec.~\ref{subsec:op}, we have shown that the  magnetic structures in Rb$_2$Mn$_3$(MoO$_4$)$_3$(OH)$_2$  has both commensurate and incommensurate components below 3.2~K, forming a complex magnetic structure with two different propagation vectors of different stars. In the language of the symmetry group, the appearance of the  second wavevector ${\bf k}_{2}=(0,~ 0,~ 0)$ means that the symmetry of the magnetic superspace group of $Pbnm.1'(00g)s0ss$ needs to be further lowered to allow the additional magnetic ordering. Effectively, the symmetry operator \{$1'\vert$000$\frac{1}{2}$\} that is associated with  the existence of a single primary incommensurate wavevector needs to be dropped. There are several maximal subgroups without this operation. The one associated with the $Pn'ma$(\#62.443) and  $Pnma.1'(0b0)s0ss$ is $Pn'ma(0b0)00s$ with an origin shift of $(0, 0, 0, 1/4)$, which agress with the experimental data very well. It can be transformed into the standard setting, $Pbn'm(00g)s00$ (62.1.9.4.m443.1), with a point group symmetry $mmm'$. This new superspace group will allow a non-zero $m=0$ term in Eq.~\ref{eq:ssg}. The symmetry constraints on the additionally ordered spin components as well as the refined results are shown in Tab.~\ref{Tab:icmSSG}, which shows that additionally ordered moment is mostly from the $c$-component of the Mn1 magnetization, supporting the previous irrep mode analysis.

As mentioned above, we have proposed a simplified magnetic structure model to describe the major components of the ordered moment. In this model, the incommensurate component and the commensurate component are mutually normal to each other, as shown in Fig.~\ref{Fig:simplemagstru}. The commensurate component consists of collinear spins on Mn1 sites with antiferromagnetically coupled moments along $c$-axis between adjacent Mn1 ions. The incommensurate component consists of cycloidally ordered magnetic moments lying in the $ab$-plane from both Mn1 and Mn2 sites. By comparing the refinements from the  irrep mode approach (using basis-vectors) and the magnetic space and superspace approach, we note that they have the same constrains and can give rise to the same results for the commensurately ordered component.  However, there are some subtle differences for the incommensurately ordered magnetic component. If we only consider the ordered moment in the $ab$ plane for the incommensurate component, the irrep mode approach shows a slightly non-collinear spin structure within each spin triangle, however the superspace approach shows a simpler structure that moments are essentially linear within each spin triangle.  In particular, the  symmetry operation of the superspace group requires that the  ordered moment in the $ab$ plane  for Mn1a and Mn1b sites are parallel to each other inside each triangle, but the irrep mode approach does not have this constraint. The two incommensurate component will be equivalent if in the irrep mode approach there was a phase difference of $0.31 \times 2\pi$ between the Mn1a and Mn1b sites and a phase difference of $0.49 \times 2\pi$ between the Mn1a and Mn2 sites. The experimental value for the Mn1a/Mn2 site ($ 0.46(2) \times 2\pi$) is close to this requirement. However, the experimental value of the Mn1a/Mn1b sites ($0.243(8) \times 2\pi$) is off without considering the symmetry constraint. Such a discrepancy between these two approaches has been seen in other systems too~\cite{rodriguez2012symmetry}.

The previously reported Rb$_2$Fe$_2$O(AsO$_4$)$_2$ shows one magnetic transition on cooling and has a commensurate single-${\bf k}$~magnetic structure at low temperature~\cite{garlea2014complex}. In contrast, Rb$_2$Mn$_3$(MoO$_4$)$_3$(OH)$_2$ shows two obvious magnetic transitions on cooling and displays a complex two-${\bf k}$~magnetic structure below 3.5~K with both a commensurate and an incommensurate components. This additional complexity may be caused by the alternating bond lengths in Rb$_2$Mn$_3$(MoO$_4$)$_3$(OH)$_2$ along the spin chain. This hypothesis requires further theoretical investigation, which is beyond the scope of the current manuscript. 

\section{Summary}
Rb$_2$Mn$_3$(MoO$_4$)$_3$(OH)$_2$ is a newly discovered bond-alternating decorated spin chain system. There are three different nearest-neighbor exchange interactions inside each individual chain. We have studied its magnetic properties and magnetic structures and found two successive magnetic phase transitions on cooling. It transitions from a paramagnetic phase into an incommensurate phase below 4.5~K. An additional antiferromagnetically ordered component arises with ${\bf k}_{2} = (0,~0,~0)$ below 3.5~K, forming  a complex magnetic structure with two different propagation vectors of different stars. On further cooling, the incommensurate wavevector undergoes a lock-in transition below 2.3~K. We have found that the  magnetic superspace group in the standard setting is $Pbnm1'(00g)s00s$ (62.1.9.4m442.2) for the single-${\bf k}$ incommensurate phase and is $Pbn'm(00g)s00$ (62.1.9.4m443.1) for the 2-${\bf k}$ magnetic structure. The experimental results show an exemplary case of complex magnetic magnetic structures that delta chain systems can host and will be the touchstone for future theoretical investigations on this intriguing bond-alternating decorated spin chain system.  

\section{acknowledgments}
Liu sincerely acknowledges Dr. Chakoumakos (ORNL) for providing suggestions to improve the manuscript, and he is also indebted to lecturers on the 5th School on Representational Analysis and Magnetic Structures (UMCP, 2015) and the Workshop on Symmetry and Superspace Approach to Modulated Crystal Structures (Oak Ridge, 2019). A portion of this research used resources at the Spallation Neutron Source, a DOE Office of Science User Facility operated by the Oak Ridge National Laboratory. Work at ORNL was partially supported by the U.S. Department of Energy, Office of Science, Basic Energy Sciences, Materials Sciences and Engineering Division. We are indebted to the NSF DMR-1808371 for support of the synthesis work.

\bibliography{main}	
\end{document}